\documentclass[letter]{aa}  
\usepackage{xcolor}

\usepackage{txfonts}
\usepackage{graphicx}
\usepackage{multirow}
\usepackage{placeins}
\usepackage{balance}
\usepackage{natbib}
\usepackage{setspace}
\usepackage{amssymb}
\usepackage[switch]{lineno}

\let\OLDthebibliography\thebibliography
\renewcommand\thebibliography[1]{
  \OLDthebibliography{#1}
  \setlength{\parskip}{0pt}
  \setlength{\itemsep}{0pt plus 0.3ex}
}

\definecolor{purple}{RGB}{102,0,204}
\definecolor{blue2}{RGB}{29,95,161}
\definecolor{orange}{RGB}{225,130,0}
\definecolor{green2}{RGB}{29,161,95}

\usepackage{hyperref,multirow}
\hypersetup{colorlinks,citecolor={blue2},linkcolor={blue2},urlcolor={blue2}}  
\newcommand{\kep}{\emph{Kepler}}
\newcommand{\prot}{$P_\text{rot}$\xspace}
\newcommand{\porb}{$P_\text{orb}$\xspace}

\newcommand{\teff}{$T_\text{eff}$\xspace}

\defcitealias{2021A&A...650A.126A}{A21b}

\begin{document}

   \title{Stellar spectral-type (mass) dependence of the dearth of close-in planets around fast-rotating stars}


\subtitle{Architecture of \emph{Kepler} confirmed single-exoplanet systems compared to star-planet evolution models}

   \author{R.~A. Garc\'\i a
          \inst{1}
          \and
          C.~Gourv\`es \inst{1}
          \and
          A.~R.~G. Santos\inst{2}
          \and
          A.~Strugarek\inst{1}
          \and
          D.~Godoy-Rivera\inst{3,4}
          \and
          S. Mathur\inst{3,4}
          \and
          V.~Delsanti\inst{5,1}
          \and
          \\ S.~N. Breton\inst{1,6}
          \and
          P.~G. Beck\inst{3,4,7}
          \and
          A.~S. Brun\inst{1}
          \and        
          S.~Mathis\inst{1}
         }
\institute{
Universit\'e Paris-Saclay, Universit\'e Paris Cit\'e, CEA, CNRS, AIM, F-91191, Gif-sur-Yvette, France\\
   \email{rgarcia@cea.fr}
    \and
    Instituto de Astrof\'isica e Ci\^encias do Espa\c{c}o, Universidade do Porto, CAUP, Rua das Estrelas, PT4150-762 Porto, Portugal 
    \and
Instituto de Astrof\'isica de Canarias (IAC), E-38205 La Laguna, Tenerife, Spain
    \and 
    Universidad de La Laguna (ULL), Departamento de Astrof\'isica, E-38206 La Laguna, Tenerife, Spain
\and
Ecole Centrale-Supelec, Universit\'e Paris-Saclay, 91190 Gif-sur-Yvette, France
\and 
INAF – Osservatorio Astrofisico di Catania, Via S. Sofia, 78, 95123 Catania, Italy
\and
Institut für Physik, Karl-Franzens Universität Graz, Universitätsplatz 5/II, NAWI Graz, 8010 Graz, Austria 
             }

   \date{Received 18 May 2023 / Accepted 20 October 2023}

 
  \abstract
   {In 2013 a dearth of close-in planets around fast-rotating host stars was found using statistical tests on \emph{Kepler} data. The addition of more \emph{Kepler} and Transiting Exoplanet Survey Satellite (TESS) systems in 2022 filled this region of the diagram of  stellar rotation period (\prot) versus the planet orbital period (\porb).
   We revisited the \prot extraction of \emph{Kepler} planet-host stars, we classify the stars by their spectral type, and we studied their \prot--\porb relations. We only used confirmed exoplanet systems to minimize biases.  
In order to learn about the physical processes at work, we used the star-planet evolution code ESPEM (French acronym for Evolution of Planetary Systems and Magnetism) to compute a realistic population synthesis of exoplanet systems and compared them with observations. Because ESPEM works with a single planet orbiting around a single main-sequence star, we limit our study to this population of \emph{Kepler} observed systems filtering out binaries, evolved stars, and multi-planets. 
   We find in both, observations and simulations, the existence of a dearth in close-in planets orbiting around fast-rotating stars, with a dependence on the stellar spectral type (F, G, and K), which is a proxy of the mass in our sample of stars. 
   There is a change in the edge of the dearth as a function of the spectral type (and mass). It moves towards  shorter \prot as temperature (and mass) increases, making the dearth look smaller.
   Realistic formation hypotheses included in the model and the proper treatment of tidal and magnetic migration are enough to qualitatively explain the dearth of hot planets around fast-rotating stars and the uncovered trend with spectral type. }

   \maketitle


%

\section{Introduction}
The architecture of observed exoplanet systems is tailored by the complex interplay between the stars and planets. It is impacted from the beginning by the structure and evolution of the protoplanetary disks and later by the tidal and magnetic interactions \citep{2018haex.bookE..24M,2018haex.bookE..25S} between all the objects in the system. In 2013, by studying 737 main-sequence \emph{Kepler} Objects of Interest \citep[KOIs,][]{2010Sci...327..977B} that had a  measured stellar rotation period (\prot),
 \citet{2013ApJ...775L..11M} uncovered the existence of a dearth of close-in planets around fast-rotating stars, as initially suggested by \citet{2009MNRAS.396.1789P}. Recently, \citet{2022ApJ...930L..23M} extended this study to 934 KOIs and 79 TESS Objects of Interest \citep[TOIs,][]{2014SPIE.9143E..20R}, and concluded that the previously mentioned dearth could be related to an observational bias. 

The engulfment of close-in planets was first proposed as a possible physical origin for such a dearth by \citet{2014ApJ...786..139T} based on tidal interactions. In their pioneering work, \citet{2014ApJ...787..131Z} carried out the first population simulations of star-planet systems, taking into account tidal interaction, which  causes  the planet to migrate inwards or outwards while making the star spin up or spin down. Based on this seminal work, \citet[][hereafter A21b]{2021A&A...650A.126A} improved the ESPEM code \citep{2019A&A...621A.124B} by including equilibrium tides, dynamical tides in the convective envelope of the host star, as well as magnetic torques from the stellar wind acting on the star and from star-planet magnetic interactions \citep[following][]{2017ApJ...847L..16S}. A similar approach was recently followed by \citet{2023MNRAS.520.3749L} to estimate the proportion of hot Jupiters that are engulfed by their star during their life on the main sequence. In addition, \citetalias{2021A&A...650A.126A} used their large grid of models to produce a synthetic population of star-planet systems that can be directly   compared with the \emph{Kepler} sample. In that initial work the imprint of the initial population of exoplanets after the disk dissipation was not considered, and it was found that the modelled stars still retained too many close-in planets compared to what was observed with \emph{Kepler}.

\citet{2013ApJ...775L..11M} and \citet{2022ApJ...930L..23M} used statistical tests to asses the existence (or lack thereof) of a dearth of close-in planets, which did not shed light on the physical processes leading to the architecture of observed exoplanet systems. In this work we follow a different approach by comparing the architecture of the observed systems to a synthetic population computed with the star-planet evolution code ESPEM, which takes into account tidal \citep{2019A&A...621A.124B} and magnetic interactions \citepalias{2021A&A...650A.126A} between a star and a single planet, from the disk-dissipation phase up to the end of the main sequence.



\begin{table*}[]\small
\caption{Stellar and exoplanet properties.}
    \centering
    \begin{tabular}{cccccccccc}
    \hline\hline
\multirow{2}{*}{KIC ID} & \multirow{2}{*}{Host ID} & $P_\text{orb}$ & $R_\text{p}$ & $P_\text{rot}$ & $T_\text{eff}$ & $\log\,g$ & [Fe/H] & Signal & Binary/Evolved\\
 & & (days) & ($R_\oplus$) & (days) & (K) & (dex) & (dex) & flag & flag \\\hline
 
757450 & Kepler-75 & 8.885 & $11.77^{+0.34}_{-0.34}$ & $19.09 \pm 1.43$ & $5301^{+111}_{-103}$ & $4.432^{+0.045}_{-0.044}$ & $0.242^{+0.130}_{-0.134}$ & -- & --  \\

1026957 & Kepler-1731 & 21.761 & 2.93 & $20.61 \pm 1.55$ & $4772^{+78}_{-73}$ & $4.595^{+0.021}_{-0.025}$ & $-0.062^{+0.102}_{-0.100}$ & -- & 1  \\

1432789 & Kepler-745 & 9.931 & $2.16^{+0.58}_{-0.22}$ & $26.43 \pm 1.41$ & $5770^{+103}_{-103}$ & $4.127^{+0.046}_{-0.055}$ & $0.188^{+0.134}_{-0.155}$ & 1 & 1  \\

1718958 & Kepler-1934 & 1.420 & 0.91 & $9.29 \pm 0.69$ & $6284^{+102}_{-103}$ & $4.467^{+0.014}_{-0.019}$ & $-0.334^{+0.101}_{-0.112}$ & -- & --  \\

1849702 & Kepler-1854 & 39.831 & 2.59 & $18.87 \pm 3.54$ & $5268^{+91}_{-83}$ & $4.402^{+0.042}_{-0.041}$ & $-0.057^{+0.119}_{-0.113}$ & 1 & 1  \\

1872821 & Kepler-1828 & 10.274 & 2.12 & $28.14 \pm 3.29$ & $5684^{+94}_{-94}$ & $4.518^{+0.021}_{-0.030}$ & $-0.117^{+0.114}_{-0.118}$ & 1 & --  \\

1873513 & Kepler-1624 & 3.290 & $5.70^{+0.39}_{-0.46}$ & $33.71 \pm 2.47$ & $3819^{+77}_{-78}$ & $4.676^{+0.013}_{-0.013}$ & $0.222^{+0.121}_{-0.123}$ & 2 & 1  \\

1996180 & Kepler-1271 & 3.026 & $1.48^{+0.23}_{-0.16}$ & $6.79 \pm 0.65$ & $6104^{+113}_{-111}$ & $4.351^{+0.034}_{-0.040}$ & $0.000^{+0.133}_{-0.133}$ & 1 & --  \\

2142522 & Kepler-1224 & 13.324 & $1.33^{+0.12}_{-0.08}$ & $10.23 \pm 1.45$ & $6319^{+110}_{-101}$ & $4.373^{+0.017}_{-0.022}$ & $0.016^{+0.110}_{-0.118}$ & 1 & --  \\
... & & & & & & & & \\\hline\hline

    \end{tabular}
    \tablefoot{Stellar parameters $T_{\rm{eff}}$, $\log g$, and [Fe/H] are from \citet{2020AJ....159..280B} or, if missing, from \citet{2017ApJS..229...30M}. Signal flag: 1 are close binary candidates, CPCB1, as described in \citep{2021ApJS..255...17S},  2 are stars for which the retrieved \prot could be half of the real one, 3 are hump \& spike candidates \citep{2023MNRAS.520..216H}. Binary/Evolved flag: 1 if it is a confirmed/candidate binary or a post-main-sequence star, or 2 if only {\it Gaia} colours are missing (for more details see Appendix~\ref{appendix:sample_selection}).\footnote{The full table is available in electronic form at the CDS via anonymous ftp to cdsarc.cds.unistra.fr (130.79.128.5) or via https://cdsarc.cds.unistra.fr/cgi-bin/qcat?J/A+A/]}}
    \label{tab:summar_table}
\end{table*}\normalsize







\section{Sample selection and stellar \prot extraction}
\label{sec:sample}

Our study is limited to observed single-planet systems to avoid introducing additional biases in the comparison between theoretical models and observations. Moreover, we use only \emph{Kepler} observations because the most complete possible \prot distribution of stars with and without detected planets is required to compute the models. 
To have the same observational biases in both samples, these periods should be assessed following the same method. This cannot be done with TESS data yet because of the difficulties in measuring rotation periods in a large sample of stars \citep[e.g.][]{2022ApJ...927..219C,2022ApJ...936..138H,2022ApJ...930....7A}, in particular for \prot longer than 10-15 days.

To elaborate the list of confirmed single-planet exosystems observed by \emph{Kepler}, we used the full list of exoplanets available at the NASA Exoplanet Archive\footnote{\url{https://exoplanetarchive.ipac.caltech.edu}} (NEA; \citealt{2013PASP..125..989A}) in March 2022. A total of 4,935 planets are registered in 3,576 systems,  2,889 of which are  single-planet exosystems. Following the methods described in Appendix~\ref{appendix:sample_selection}, we obtained 1,967 confirmed planet systems observed by \kep,\ of which 1,476 are single-planet systems (based on the catalogues used here).



Two sets of light curves (LCs) were used in this work to look for \prot: Pre-search Data Conditioning -- Maximum A Posteriori \citep[PDC-MAP,][]{2010ApJ...713L..87J,2012PASP..124..985S,2012PASP..124.1000S,ThompsonRel21} and our custom KEPSEISMIC\footnote{\url{https://archive.stsci.edu/prepds/kepseismic/}}  \citep{2011MNRAS.414L...6G}. More details on the corrections and the data preparation are given in Appendix~\ref{appendix:data}.

The values of \prot  were obtained using the automatic selection procedure described in \citet{2021ApJS..255...17S} coupled to the machine learning algorithm ROOSTER \citep{2021A&A...647A.125B}. Moreover, all the stars (with and without a retrieved \prot) were visually inspected using the three new folded-KEPSEIMIC and the PDC-MAP LCs. 
In Table~\ref{tab:summar_table} we provide the list of 796 stars with \prot after the visual checks. 

For the moment, ESPEM is only  optimized for systems in which the central star is single and on the main sequence. Hence, it is necessary to remove post-main-sequence stars as well as binary systems. To do so, we used the astrometric and photometric data from releases EDR3 and DR3 of the {\it Gaia} mission \citep{2021A&A...649A...1G,2023A&A...674A...1G}. This allowed us to remove potentially evolved exoplanet hosts, as well as different categories of binary systems. We give further details of the selection cuts in Appendix \ref{appendix:sample_selection}. After applying these cuts, we ended up with 576 confirmed single-planet-host main-sequence solar-like stars (CSPHMSS) with a reliable rotation period. To perform meaningful comparisons, we   applied the same selection cuts to the latest \emph{Kepler} rotation catalogue  \citep{2019ApJS..244...21S,2021ApJS..255...17S} to remove post-main-sequence stars and binary systems, which  yielded our reference \emph{Kepler} sample (RKS, see Appendix~\ref{appendix:ref_kep}) for the remainder of the paper.




\section{Observed \porb versus \prot distributions}
\label{sec:Porb_Prot_distributions}

\begin{figure}[!hb]
   \centering
   \includegraphics[width=0.425\textwidth]{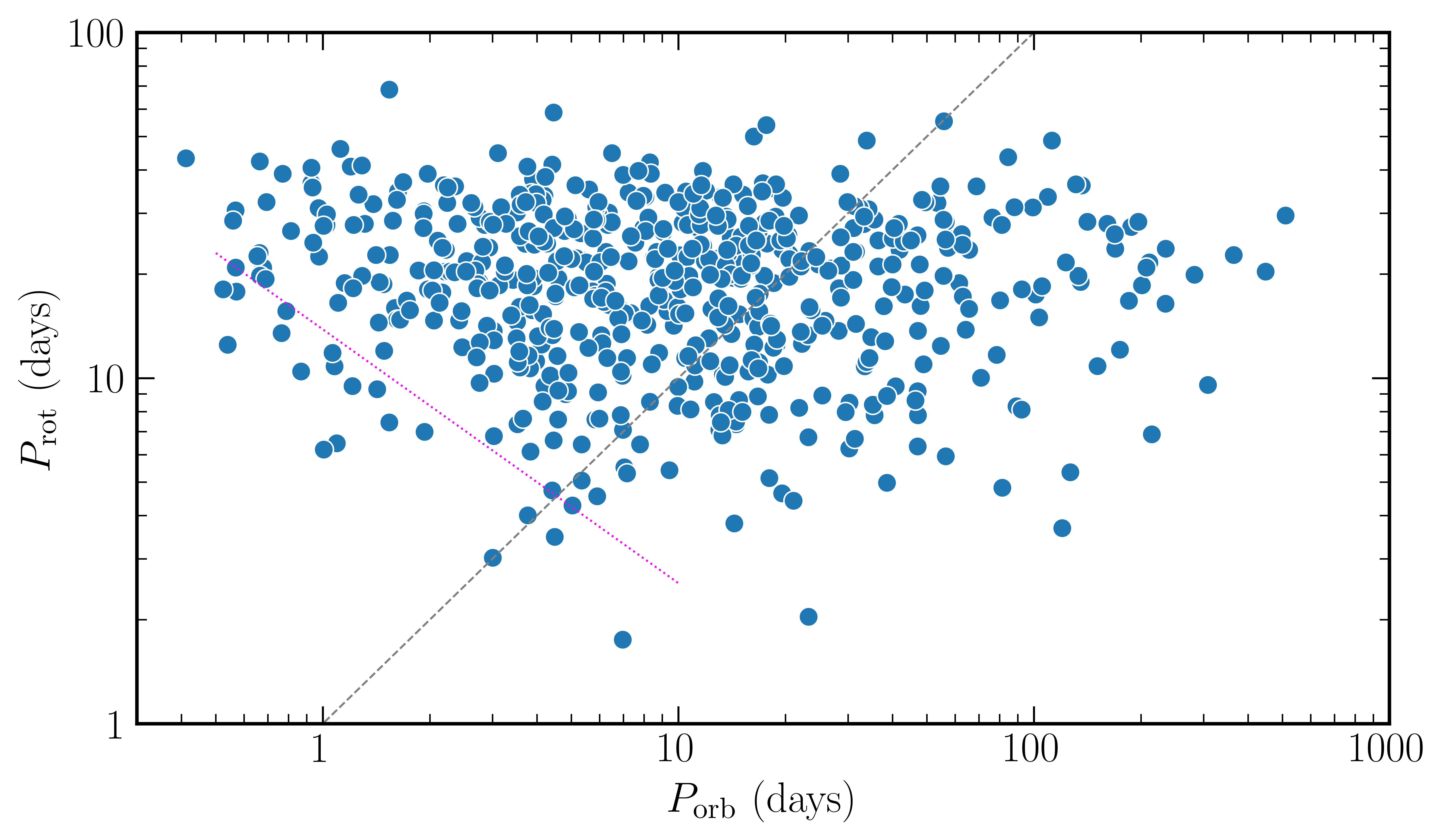}
   \caption{$P_\text{rot}$ as a function of $P_\text{orb}$ for the  CSPHMSS sample. The grey dashed line corresponds to the 1:1 line (synchronization). The magenta dotted line is the fit to the lower envelope of points obtained by \citet{2013ApJ...775L..11M}.  }
              \label{Fig1}%
    \end{figure}

In this section we investigate the \prot\ and planet orbital period (\porb) distributions and their relationship. Figure \ref{Fig1} shows the rotation period as a function of the orbital period for 576 CSPHMSS of all spectral types, similarly to Fig.~2 in \citet{2013ApJ...775L..11M}. 
\begin{figure*}[ht]
    \centering
    \includegraphics[width=0.76\textwidth]{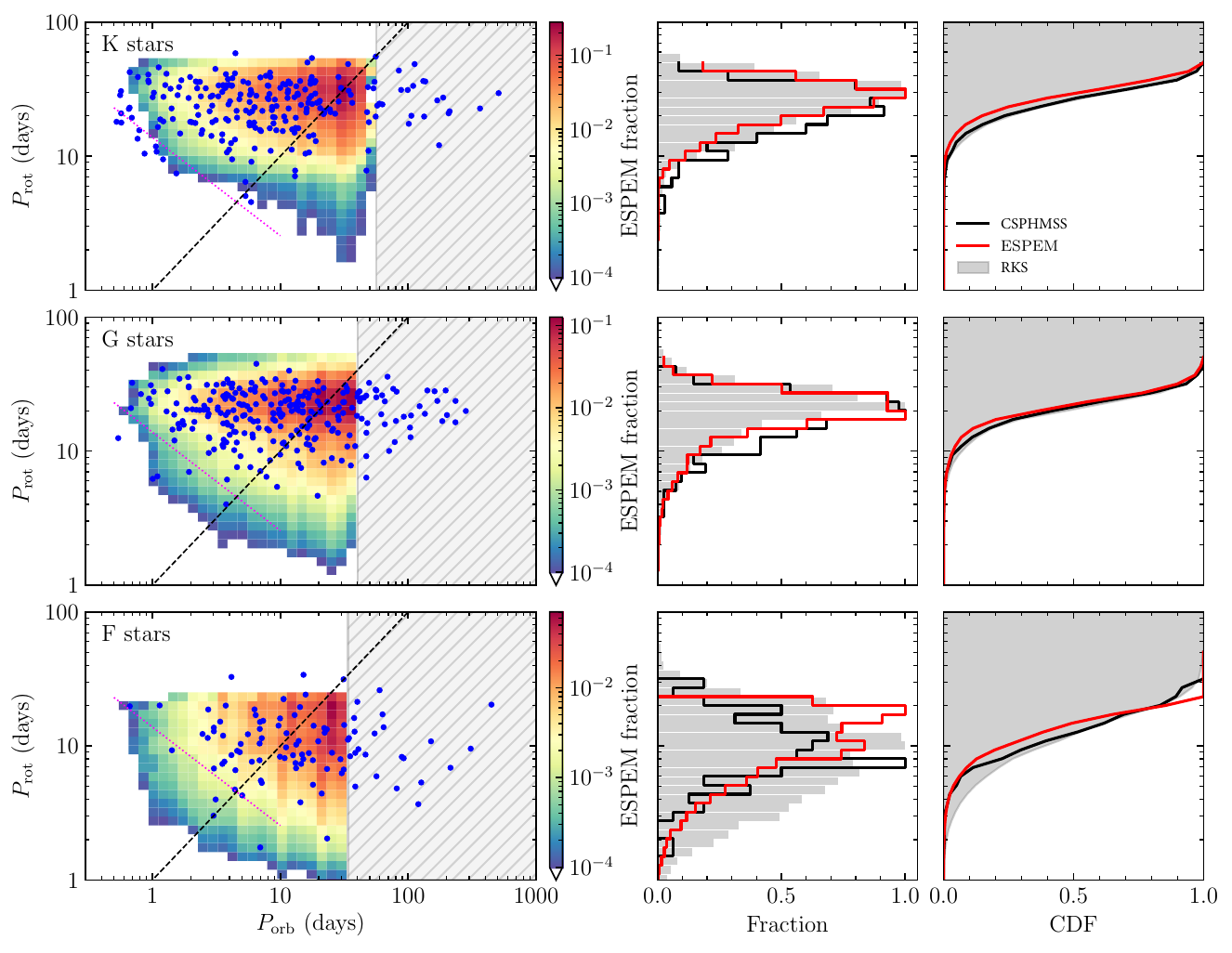}
    \caption{\prot\ vs \porb\ per spectral type (K-, G-, and F-type stars from top to bottom,  left panels). The coloured regions depict the distribution of star-planet occurrences computed with ESPEM, while the blue dots are the CSPHMSS. The grey shaded region indicates the parameter space not covered by the simulation. The dotted and dashed lines are the same as in Fig.~\ref{Fig1}. The middle and right panels correspond to the \prot histograms and CDFs, respectively, for the three distribution of stars: CSPHMSS (black), ESPEM (red), and RKS (grey).}
    \label{fig:Prot_Porb_spectraltype}
\end{figure*}

While the overall picture is similar in the two analyses, our samples differ (see a detailed comparison with this and other previous works in Appendix~\ref{appendix:comp}). We note that there are a few more stars below the edge of the dearth of close-in planets region and fewer fast rotators. Although there are  more fast rotators in \citet{2013ApJ...775L..11M}, the \prot distributions become similar when the same selection criteria are applied (see Appendix~\ref{appendix:comp}).

Exoplanet occurrence rates on \emph{Kepler} fast-rotating main-sequence stars depend on the  spectral types \citep[F dwarfs spin faster than G and K dwarfs,][]{2019ApJS..244...21S, 2021ApJS..255...17S}. Hence, we also separated our exoplanet sample by spectral types (K to F, see blue dots and black lines in Fig.~\ref{fig:Prot_Porb_spectraltype}). 
M dwarfs were excluded from this study because there are only five, which is not enough to draw any reliable conclusions. 

Because a large fraction of the stars in the \emph{Kepler} field have near solar metallicity \citep{2014ApJ...789L...3D} and because our planet hosts are all main-sequence stars, the spectral type (defined using $T_{\rm eff}$) is a very good proxy of stellar mass.  Hence, the ranges of masses \citep[from][]{2020AJ....159..280B} corresponding to our cuts in spectral type are for K, G, and F: $0.436 \lesssim M \lesssim 0.896$, $0.769 \lesssim M \lesssim  1.162$, and $ M \gtrsim 1.015$ M$_\odot$, respectively, where the  $\odot$ quantities refer to solar values. There is some overlap at the edges of the spectral types as the relative mass uncertainties are two to three times larger than those on $T_{\rm eff}$ (from $\sim$2 to $\sim$9~$\%$).


Qualitatively speaking, stars in the \prot--\porb diagram follow several trends with $T_{\rm eff}$ and mass (see left panels in Fig.~\ref{fig:Prot_Porb_spectraltype}). First, there are more fast-rotating F dwarfs   than G and K dwarfs at all \porb, a behaviour similar to that in the  RKS. Second, there are more slow-rotating K dwarfs with close-in planets than G and F dwarfs. For F dwarfs there is a general lack of close-in planets at all \prot, except for three stars below \porb = 2 days.

There is also a larger number of synchronized systems for G stars: adopting a \prot/\porb\ tolerance of 5\%, we find 23 synchronized systems (9\% of the sample) for G stars, and 12 (5\%) and 4 (4\%) for K and F stars, respectively. Although captivating, their investigation is beyond the scope of this work.

To examine the distribution of CSPHMSS  against the RKS, we generated the histogram and the cumulative distribution function (CDF) by spectral types (Fig.~\ref{fig:Prot_Porb_spectraltype} middle and right panels, respectively). For the K dwarfs the two samples follow  similar distributions. We performed the Kolmogorov-Smirnov  \citep[KS,][]{Kolmogorov1933,Smirnov1939} test to quantify their differences. We find a KS statistic of 0.07, with a p-value of 0.29, 
suggesting that for the K dwarfs  CSPHMSS and RKS are both drawn from the same underlying distribution. In the G-dwarf CDF we see a lack of fast rotators in the CSPHMSS below $\sim$12 days (0.08 KS statistic with a p-value of 0.11 supporting the null hypothesis). Finally, for the F dwarfs we find a clear deficit of CSPHMSS rotating faster than 7 days, while there is an excess of \prot between 7 and 15 days, with a peak at 7-8 days as shown in the histogram. For this sample, the KS statistics increases significantly (0.27), with a p-value of $1.24\times 10^{-6}$, rejecting the null hypothesis.

\section{Synthetic population of rotating stars and hot exoplanets with ESPEM}
\label{sec:ESPEMpop}


We assessed the plausible physical scenarios behind the distribution of star-planet systems via the ESPEM code. 
ESPEM solves a budget of momentum between the angular momentum in the circular orbital motion of a planet, the angular momentum in the convective envelope of a solar-like star, and the angular momentum within the radiative interior of the same star. It takes into account coupling between the two zones of the star \citep{1991ApJ...376..204M,2015A&A...577A..98G},   stellar wind torque on the star \citep{2015ApJ...799L..23M}, stellar evolution during the pre-main sequence and the main sequence \citep{2016A&A...587A.105A}, and exchanges between the planet and the star due to tidal \citep{2015A&A...580L...3M,2013MNRAS.429..613O,2016CeMDA.126..275B} and magnetic interactions \citep{2016ApJ...833..140S,2017ApJ...847L..16S}. The complete description of the model can be found in \citetalias{2021A&A...650A.126A}; in this   work a methodology was developed to build a synthetic population of star-planet systems comparable with the \emph{Kepler} sample of \citet{2013ApJ...775L..11M}. We revisit  this work here, and compare such a synthetic population with the CSPHMMS and RKS samples.

To build a synthetic population, we started by producing three sets of 8,000 ESPEM models. Each set includes 40 different initial semi-major axes between 0.005 and 0.2 AU, five different initial rotation periods between 1 and 10 days, five planetary masses between 0.5 $M_{\rm Earth}$ and 5 $M_{\rm Jupiter}$, and eight stellar masses between 0.5 and 1.2 M$_\odot$. Each set of 8,000 models includes its own physical ingredients: set~1 considers only tidal interaction; set~2 tidal and magnetic interactions with a planetary magnetic field strength of 1 G; and set~3 is like set~2, but with the planetary field of 10 G (see \citetalias{2021A&A...650A.126A} for more details). Within these 8,000 models, we select instantaneous stellar evolution epochs for a given stellar mass to mimic the distribution of the RKS. The main difference between the three sets lies in the population of  low-mass planets \citepalias{2021A&A...650A.126A}, where the magnetic torques   dominate their migration and affect their distribution. Nevertheless, these planets hold low angular momentum in their orbit, and therefore their net effect on the population of stars is negligible. In this work we primarily focus on the population of stars with close-in planets, hence in what follows we only consider set~3 and defer a detailed comparison of planetary populations between ESPEM to a future publication.


The comparison between ESPEM  and \emph{Kepler} results from \citet{2013ApJ...775L..11M} made by \citetalias{2021A&A...650A.126A} showed   good qualitative agreement, but the synthetic populations exhibited an excess of close-in planets. Several reasons could be at the origin of this discrepancy: an enhanced magnetic interaction of close-in planets harbour larger surface magnetic fields than 10 G \citep{2017ApJ...849L..12Y,2021ApJ...908...77H}; additional star-planet angular momentum exchanges, for example  due to  tidal interactions with the radiative core of the star \citep{2021A&A...651A...3A};   a bias due to the selection of the initial semi-major axes used in the ESPEM set that should correspond to a population of planets right after the dissipation of the disk. Here we consider the last hypothesis and remove from the ESPEM population all the planets that were initiated below the inner radius of the dead zone ($R_{\rm dz}$) of the stellar disk. We provide an analytical derivation of $R_{\rm dz}$ in Appendix \ref{app:rdz}, as well as a formulation of $R_{\rm dz}$ as a function of $M_\star$. It should be noted that we also consider the transit detection probability when building up our synthetic population of rotating stars with close-in exoplanets (for more details, see Section 5 of \citetalias{2021A&A...650A.126A}).


We show in the left panels of Fig. \ref{fig:Prot_Porb_spectraltype} the ESPEM probability of the occurence of star-planet systems   as a function of \prot and \porb in logarithmic scale colour map. The ESPEM population is separated into the three different spectral types   we focus on in this work (K, G, and F dwarfs). The predicted \prot ranges from 0.6 to 50 days and the orbital period ranges from 0.3 to 50 days. In ESPEM we do not consider planets beyond a given \prot, as indicated by the dashed grey area that labels the parameter space not covered by the simulations. 

The simulated populations also exhibit trends with spectral type, as shown in Fig.~\ref{fig:Prot_Porb_spectraltype}: a larger fraction of fast-rotating F dwarfs that progressively decreases for G and K dwarfs, and a lower edge of the dearth well aligned with the lower envelope of points deduced by \citet{2013ApJ...775L..11M} (magenta dotted lines in left panels of Fig.~\ref{fig:Prot_Porb_spectraltype}). This is particularly striking for K-dwarf systems with \porb between 1 and 40 days.  

Compared to the initial work of \citetalias{2021A&A...650A.126A}, we obtain in this work fewer close-in planets, thanks to a refined selection of the initial star-planet population since we do not consider here planets with orbits closer to the inner $R_{\rm dz}$  of the protoplanetary disk. This clearly  shows the importance of taking into account planetary formation processes. In addition, we   also explored the effect of the chosen mass-loss rates and stellar magnetic-field laws on our population synthesis. Following \citet{2020A&A...635A.170A}, we   considered two extreme sets of mass-loss rate and magnetic field laws, covering the range of possibilities still compatible with all the observational constraints at hand to date. This extended study is shown in Appendix \ref{app:laws}, and we note that none of the conclusions in this work are strongly affected by choosing one specific set of evolutionary laws.

In all cases the dearth originates from angular momentum exchange between the rotating star and the orbit of the planet by tidal and magnetic interactions, making hot exoplanets around fast-rotating stars migrate efficiently inwards (if \porb is shorter than \prot, {i.e.} left of the dashed black line) and outwards (if \porb is longer than \prot, {i.e.} right of the dashed black line). Interestingly, the population of planets on the shortest period orbits around $P_{\rm rot}=30$ days in the top left panel stems from the migration of these planets from an initial longer orbital period. We find that this population is shaped by both tidal and magnetic effects, since the ESPEM set 1 only predicts a scarce amount of planets in this region (see Appendix \ref{sec:ESPEM_tideonly}). Finally, the simulated \prot distribution of planet hosts shifts towards slower \prot compared to the RKS population, confirming the results obtained by \citet{2022MNRAS.513.2057S}.

\section{Discussion and conclusions}
\label{sec:conclusions}







Comparing the CSPHMSS to the ESPEM predictions, we find a general good agreement (see Fig.~\ref{fig:Prot_Porb_spectraltype}). The high-density regions (red area in the left panels) closely match   the locations with the higher number of systems detected by \emph{Kepler}, particularly for the G-type stars (0.769 $\lesssim M \lesssim$  1.162  M$_\odot$). It is important to note that there is a factor of  100 difference between the red  and  green colours in the density scale. Thus, it is normal that only a few observed systems populate the lower density regions. For cooler K-type stars (top panel, 0.436 $\lesssim M \lesssim$ 0.896 M$_\odot$),  a greater number of slow-rotating stars hosting close-in planets are observed compared to the simulation. Therefore, in its current configuration, planets do not migrate fast enough in ESPEM to repopulate this region of the diagram. A possible solution could be to increase the stellar magnetic field for this spectral type, which  would strengthen the magnetic torques (see the comparison with Fig.~\ref{fig:Prot_Porb_spectraltype_tidleonly} made including only tidal interactions),
or to take into account the dissipation of tidal waves within the radiative core of stars \citep{2021A&A...651A...3A}. 

Since the number of observed F-type stars (M $\gtrsim$ 1.015 M$_\odot$) is small, it prevents us from properly interpreting how well the simulation behaves close to low-density regions. The comparison of the histograms (middle panels in Fig.~\ref{fig:Prot_Porb_spectraltype}) shows that the shapes of the three distributions (RKS, ESPEM, and CSPHMSS) are very similar,  but there is a progressive reduction in the overall number of stars for each \prot bin. However, there are two noticeable differences at \prot of around 8 and 15-20 days, where there is an excess of CSPHMSS and of simulated systems, respectively. In this last case the discrepancy may be due to an overestimation of the stellar wind-break effect \citep[whose rotation evolution is known to be difficult to capture accurately; see \text{e.g.}{}][]{2019A&A...631A..77A}. This will be investigated in a future study.

Up to \prot\,$\sim$\,7 days, the three distributions of the G-type stars show the same slope in the histograms. An excess of CSPHMSS is clearly visible up to \prot\,$\sim$\,15 days. For K-type stars the differences are minimal, with just an overall excess of observed systems compared to the simulated systems. 

Finally, CDFs are shown in Fig.~\ref{fig:Prot_Porb_spectraltype} (right column). In all of them  a steep slope is observed, indicating a more concentrated distribution of exoplanets around stars with a longer \prot for all   three samples. The ESPEM CDFs (red lines) reflect that the model reproduces relatively well the K- and G-dwarf  populations (black lines), and shows a slight overdensity of fast-rotating F stars.      

In conclusion, ESPEM models predict a dearth, in agreement with \emph{Kepler} CSPHMSS, and with a clear dependence on the stellar spectral type. More observations will be necessary to better constrain the dearth and the mechanisms at play to define the architecture of exoplanet systems.

\begin{acknowledgements}
The authors of this paper acknowledge James Davenport for providing the KIC2TIC as an open source code at github. We also thank David V. Martin for providing an updated list of known circumbinary exoplanet hosts and C. Le Poncin-Lafitte for his contribution to ESPEM. A.S.B, S.N.B., R.A.G., A.S., and St.M. acknowledge the support from the PLATO Centre National D'{\'{E}}tudes Spatiales grant. A.S. acknowledges funding from the European Union’s Horizon-2020 research and innovation program (grant agreement no. 776403 ExoplANETS-A). A.S. and St.M. acknowledge funding from the Programme National de Plan\'etologie (PNP). A.R.G.S. acknowledges the support from the FCT through national funds and FEDER through COMPETE2020 (UIDB/04434/2020 \& UIDP/04434/2020)
and the support from the FCT through the work contract No. 2020.02480.CEECIND/CP1631/CT0001. Sa.M. ~acknowledges support from the Spanish Ministry of Science and Innovation (MICINN) with the Ram\'on y Cajal fellowship no.~RYC-2015-17697 and through AEI under the Severo Ochoa Centres of Excellence Programme 2020--2023 (CEX2019-000920-S). 
Sa.M. and D.G.R. acknowledge support from the  Spanish Ministry of Science and Innovation (MICINN) with the grant no.~PID2019-107187GB-I00. PGB acknowledges support by the Spanish Ministry of Science and Innovation with the 
\textit{Ram{\'o}n\,y\,Cajal} fellowship number RYC-2021-033137-I and the number MRR4032204. S.N.B acknowledges support from PLATO ASI-INAF agreement n.~2015-019-R.1-2018. P.G.B. acknowledges the financial support by \textit{NAWI\,Graz}. The work presented here was partially supported by the NASA grant NNX17AF27G.
This paper includes data collected by the \kep\, mission and obtained from the MAST data archive at the Space Telescope Science Institute (STScI). Funding for the \kep\ mission is provided by the NASA Science Mission Directorate. STScI is operated by the Association of Universities for Research in Astronomy, Inc., under NASA contract NAS 5–26555. This research has made use of the NASA Exoplanet Archive, which is operated by the California Institute of Technology, under contract with the National Aeronautics and Space Administration under the Exoplanet Exploration Program. This research was supported in part by the National Science Foundation under Grant No. NSF PHY-1748958. The authors acknowledge the participants of the ``Probes of Transport in Stars (transtar21)''  KITP Program for the usefull discussions and comments about this work.

\textit{Software:} AstroPy \citep{astropy:2013,astropy:2018}, KADACS \citep{2011MNRAS.414L...6G}, ROOSTER \citep{2021A&A...647A.125B}, Matplotlib \citep{matplotlib}, NumPy \citep{numpy}, SciPy \citep{scipy}, Seaborn \citep{seaborn}, pandas \citep{mckinney-proc-scipy-2010_pandas,reback2020pandas},  ESPEM \citep{2021A&A...650A.126A}.

\end{acknowledgements}

%
%
\bibliographystyle{aa} 
\bibliography{BIBLIO.bib}

\begin{appendix}

\section{Sample selection}
\label{appendix:sample_selection}
At NEA we select  the table containing only confirmed exoplanet systems. Because this table contains TESS Input Catalog identifiers \citep[TIC identifiers,][]{2019AJ....158..138S},  we cross-match  the TICs with the \kep\ Input Catalog identifiers \citep[KIC identifiers,][]{2011AJ....142..112B} using the KIC2TIC tool\footnote{\url{https://github.com/jradavenport/kic2tic}} to know the systems observed by \emph{Kepler}.
Then we select  the systems with the column \texttt{sy\_pnum}= 1 to ensure that they are single exoplanet systems. To remove already known binaries, we select those systems with the column \texttt{sy\_snum}=1 and \texttt{cb\_flag}=0 (to also remove circumbinary planets).

As mentioned in Sections \ref{sec:sample} and \ref{sec:Porb_Prot_distributions}, for both the exoplanet host and RKS samples, our analysis is focused on single main-sequence stars. For this purpose, we   removed evolved stars and binary systems taking full advantage of the state-of-the-art {\it Gaia} data, using an identical procedure for both samples. Here we provide a summary of these selections, and refer interested readers to Godoy-Rivera et al. (in preparation) for further details.

\begin{figure}[h]
    \centering
    \includegraphics[width=\hsize]{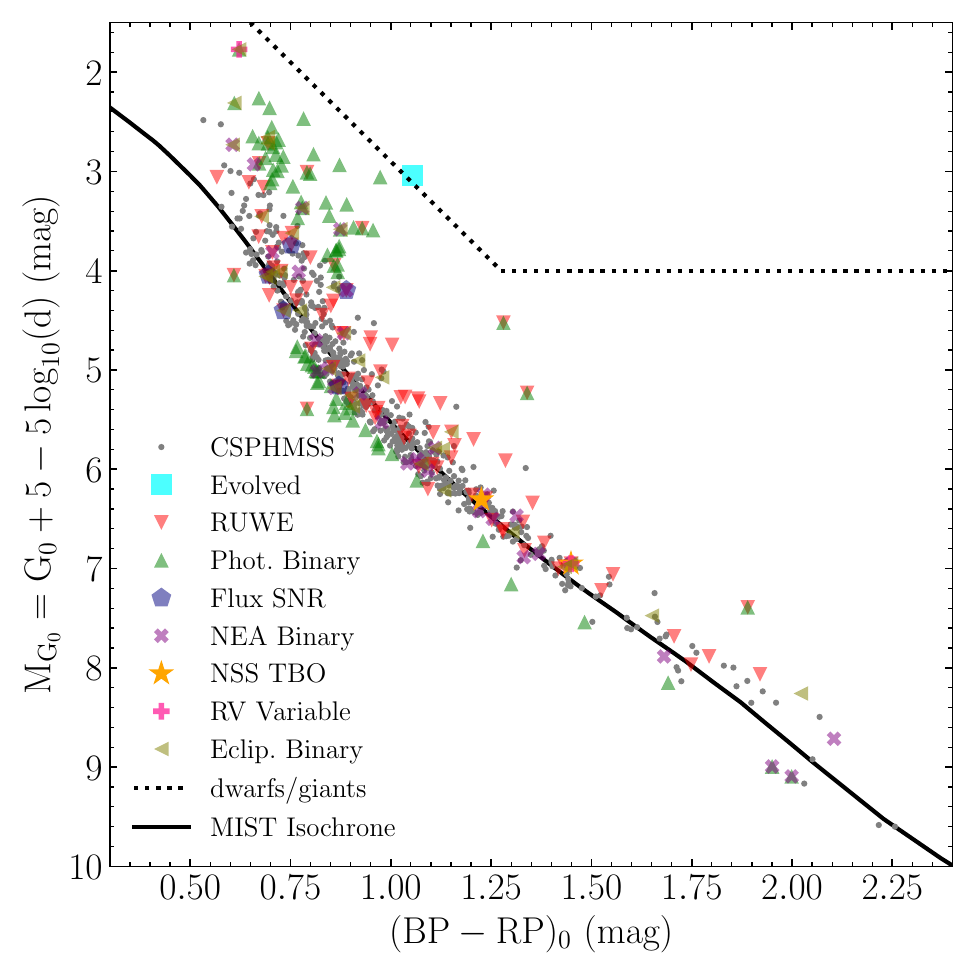}
    \caption{Absolute and de-reddened CMD of the sample of confirmed single-planet-host stars with measured \prot. The grey points represent the stars that survived all the selection cuts, i.e. the 576 CSPHMSS. The different categories described in Appendix \ref{appendix:sample_selection} are highlighted according to the legend. The MIST isochrone of the reference population is shown as the solid black line. The dotted line represents the separation between dwarf and evolved stars.}
    \label{fig:CMD_selection}
\end{figure}

As some of our cuts depend directly on the position on the colour-magnitude diagram (CMD), we first apply a quality criterion and we flag stars that lack {\it Gaia} magnitudes or that have a flux signal-to-noise ratio (S/N) in any of the three {\it Gaia} bands ($G$, $BP$, or $RP$) of \texttt{phot\_mean\_flux\_over\_error} $\leq 100$. We used the distance information from the {\it Gaia} EDR3 catalogue by \citet{2021AJ....161..147B}. To de-redden the photometry, we used the {\it Gaia} DR3 \texttt{gspphot} values when available, and otherwise used the \citet{2019ApJ...887...93G} extinction values or the  Total Galactic Extinction (TGE; \citealt{2023A&A...674A..31D})   values (queried via the \texttt{dustmaps} package;  \citealt{2018JOSS....3..695M}) following the approach of \citet{2021ApJS..257...46G}. With these values  the  stars are placed on the absolute and de-reddened CMD. We illustrate this for the confirmed single-planet-host star sample with measured \prot in Figure \ref{fig:CMD_selection}, showing the stars with enough data to be placed on the diagram (i.e. with photometric, distance, and extinction information).

In this way evolved stars are identified by applying a cut in the CMD. We define a line that distinguishes between dwarfs and evolved stars based on the whole \emph{Kepler} sample, and we illustrate it as the black dotted line in Figure \ref{fig:CMD_selection}. Stars that fall inside this region (i.e. redder colours and more luminous absolute magnitudes than the dotted line) are flagged as evolved. Binary stars are placed in one of  six categories: 1) RUWE binaries (\citealt{2021A&A...649A...2L}), meaning stars with high astrometric noise in the  {\it Gaia} solution, are identified as stars with Renormalized Unit Weighted Error (RUWE) $> 1.2$ \citep{2020AJ....159..280B}; 2) photometric binaries are identified on the CMD as dwarf stars above 0.9 mag or below 0.3 mag of the reference population MIST isochrone \citep{2016ApJS..222....8D,2016ApJ...823..102C,2011ApJS..192....3P,2013ApJS..208....4P,2015ApJS..220...15P}, which we take to be a 200 Myr and [Fe/H]=+0.25 dex model following \citet{2022ApJ...930L..23M} and \citet{2021ApJ...913...70G}; 3) stars that are found in the cross-match by \cite{2023arXiv230710812B} with the {\it Gaia} DR3 NSS TBO (Non-Single-Star Two-Body-Orbit; \citealt{2023A&A...674A..34G}) table; 4) radial velocity (RV) variable stars are identified from the {\it Gaia} RV measurements following \citet{2023A&A...674A...5K}; 5) NEA binaries are identified as   multiple systems according to the NEA database (\texttt{sy\_snum}>1); and  6) eclipsing binaries are identified from the third revision of the \emph{Kepler} Eclipsing Binary Catalog.\footnote{\url{http://keplerebs.villanova.edu/}}

For purposes of completeness, we also performed a check to identify potential circumbinary stars. We cross-matched our targets with the list of known circumbinary exoplanet hosts (e.g. \citealt{2018haex.bookE.156M}), and found seven of them in our KOI sample: KIC 5473556 (Kepler-1647), KIC 6504534 (Kepler-1661), KIC 6762829 (Kepler-38), KIC 8572936 (Kepler-34), KIC 9632895 (Kepler-453), KIC 12351927 (Kepler-413), and KIC 12644769 (Kepler-16). We note that all  these targets also correspond to NEA-identified binaries, and hence no further binary category was needed to flag them.

Regarding the sample of confirmed single-planet-host stars with measured \prot, we illustrate the CMD location of the above-mentioned categories in Figure \ref{fig:CMD_selection}, and report their respective numbers in Table \ref{tab:newcuts}. After applying the selection criteria, we are left with 576 stars, which compose the CSPHMSS sample analysed throughout this paper. An analogous version of this analysis for the full \emph{Kepler} sample is shown in Godoy-Rivera et al. (in preparation).

\begin{table}[h]
    \caption{Categories and their respective numbers of evolved and confirmed non-single planet-host stars.}
    \centering
    \begin{tabular}{l|r}
        Criteria & N stars\\\hline\hline
        Flux S/N & 11 \\
        CMD Evolved & 1 \\
        Binary RUWE & 111\\
        Binary CMD Photometric & 83 \\
        Binary NSS TBO & 2 \\
        Binary RV variable & 2 \\
        Binary NEA & 32 \\
        Binary Eclipsing & 34 \\
        \hline
        CSPHMSS & 576 \\
       \hline
    \end{tabular}
    \tablefoot{The categories are not mutually exclusive (i.e. a given star can belong to one or more categories simultaneously).}
    \label{tab:newcuts}
\end{table}

\begin{figure}[h]
    \centering
    \includegraphics[width=\hsize]{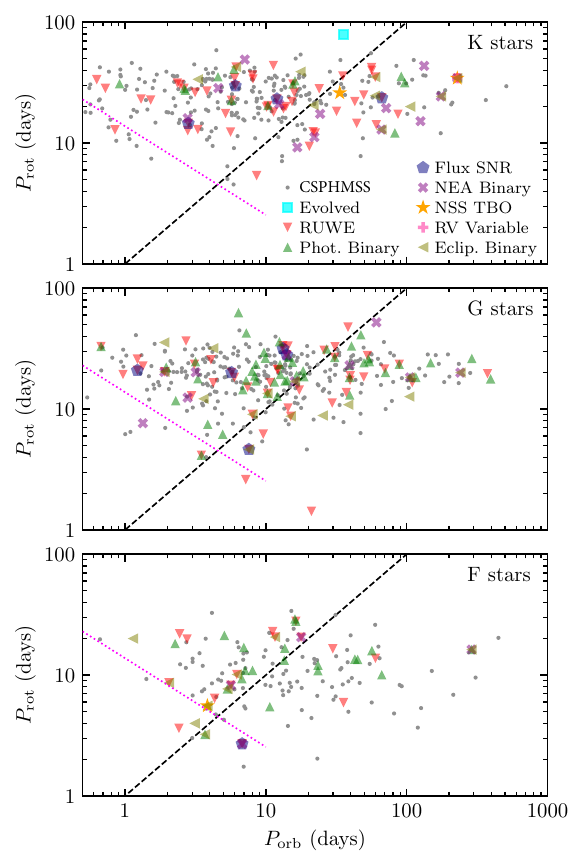}
    \caption{Same as  Fig.~\ref{Fig1}, but separated by spectral type and highlighting the stars that are not main-sequence and single stars.}
    \label{fig:Prot_Porb_cuts}
\end{figure}

Figure~\ref{fig:Prot_Porb_cuts} shows the stellar rotation period as a function of the planet orbital period, highlighting the stars that were considered to be in multiple systems or evolved according to the criteria above. Most of the removed stars are above the dearth line defined by \citet{2013ApJ...775L..11M}, but a few of these targets are located in the dearth.

\section{\emph{Kepler} light curve preparation}
\label{appendix:data}

The PDC-MAP and KEPSEISMIC light curves are from the Mikulski Archive for Space Telescopes (MAST) archive, and in both sets we follow the KADACS \citep[\emph{Kepler}Asteroseimic data analysis and calibration Software;][]{2011MNRAS.414L...6G} procedures to remove outliers, correct any jumps and drifts, and stitch together the \kep\ quarters. Because in this work we are studying stars with transiting planets, the LC segments in which the planets are transiting are removed and all the gaps are interpolated using in-painting techniques based on a multi-scale discrete cosine transform \citep{2014A&A...568A..10G,2015A&A...574A..18P,2021A&A...648A.113B}. To remove the long-period trends (mostly instrumental), the KEPSEISMIC light curves are then filtered at low frequency with three different high-pass triangular filters of 20, 55, and 80 days, as   done in previous analyses of the surface rotation using our rotation pipeline \citep[e.g.][]{2014A&A...572A..34G,2016MNRAS.456..119C,2019ApJS..244...21S,2021ApJS..255...17S}. However, we   also produced three additional LCs per star after doing a new correction in KADACS prior to the filtering, in order to minimize the impact of the \emph{Kepler} annual modulation: $P_\text{\emph{Kepler}} \sim 372.5$ days. For each star the auto-correlation function (ACF) of the light curve is performed, and we look for the highest peak around $\pm$ 30 days of $P_\text{\emph{Kepler}}$. Then, the LC is phase-folded with this periodicity and fitted.  The corrected light curve is obtained by subtracting this fitted curve linearly scaled to the data in each segment of length $P_\text{\emph{Kepler}}$. We   verified that the new folded LCs are in general flatter and less perturbed when analysing the longer filters (55 and 80 days), helping the inference of longer \prot. Finally, we filter the PDC-MAP LCs at 54 days. However, it is important to mention that, in the original PDC-MAP LCs, each quarter could be intrinsically filtered (or not) independently from the rest of them with a cut-off period between 3 and 20 days. Therefore, in several stars some quarters could be filtered and others not. This could lead to spurious periodic signals in the PDC-MAP LCs.

\section{Defining the reference \emph{Kepler} sample}
\label{appendix:ref_kep}

To make a fair comparison of the CSPHMSS sample with the broad \kep\ sample of stars with known rotation periods, we need to build our reference \kep\ sample (RKS) by removing known KOIs and by applying the same selection criteria that was considered for the CSPHMSS (Appendix~\ref{appendix:sample_selection}). The starting sample is that of \citet{2019ApJS..244...21S, 2021ApJS..255...17S}, which contains subgiant stars by design, but this work focuses on main-sequence stars. After removing the confirmed planet hosts and the stars that did not pass the criteria from the  Appendix B, the reference sample for this study includes 34,265 dwarfs: 316 M dwarfs ($T_\text{eff}<3,700$ K); 12,081 K dwarfs ($3,700\leq T_\text{eff}<5,200$ K); 14,705 G dwarfs ($5,200\leq T_\text{eff}<6,000$ K); and 7,163 F dwarfs ($T_\text{eff}>6,000$ K).

\begin{figure}
    \centering
    \includegraphics[width=\hsize]{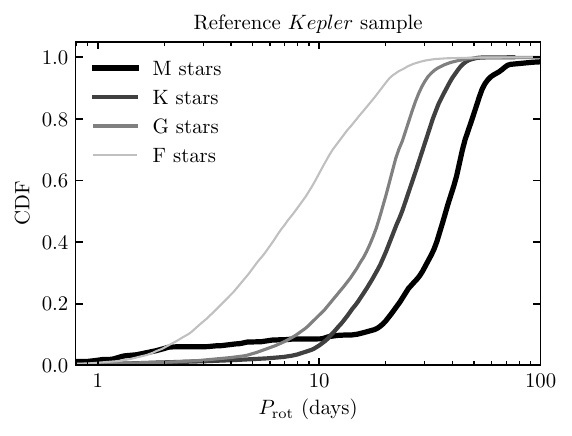}
    \caption{Cumulative distribution function for \prot\ per RKS spectral type,   indicated by the thickness and shade of the lines (see legend).}
    \label{fig:Prot_dist-Teff}
\end{figure}

As shown previously \citep[e.g.][]{2013ApJ...775L..11M,2021ApJS..255...17S}, the rotation period of \kep\ stars generally decreases with \teff. Figure~\ref{fig:Prot_dist-Teff} illustrates this trend for the reference sample considered in this work by comparing the \prot\ distribution of each spectral type. This \teff dependence is relevant in the context of the dearth of close-in planets orbiting fast-rotating stars.

\FloatBarrier

\section{Comparison with previous KOI studies}
\label{appendix:comp}

In this section we compare our rotation estimates and rotation distribution with those in the previous KOI catalogues. 
We consider the KOIs and respective \prot\ from \citet[][hereafter WB2013]{2013MNRAS.436.1883W}, \citet[][hereafter MMA2013]{2013ApJ...775L..11M}, \citet[][hereafter MPM2015]{2015ApJ...801....3M}, and \citet[][hereafter MOG2022]{2022ApJ...930L..23M}. WB2013 analysed the KOI long-cadence data corresponding to \kep\ Quarter (Q) 9. Performing a periodogram analysis, the authors reported periods for 954 KOIs. MMA2013 retrieved the rotation periods of 737 KOIs by computing the ACF of the light curves from Q3 to Q14. Similarly to MMA2013, MPM2015 analysed Q3-Q14 data using the ACF, being able to report reliable periods for 993 KOIs in the main sequence. All these works adopted PDC-MAP data products. In MOG2022, the authors adopted the \prot\ values for KOIs reported by the previous works: WB2013, MMA2013, and MPM2015. Additionally, MOG2022 applied selection criteria to remove possibly evolved and binary stars (see Appendix~\ref{appendix:sample_selection}).

In this work we revisit the \kep\ long-cadence data, but making use of the full length of the observations, from Q0 to Q17, when available. Although we also use PDC-MAP LCs for the visual inspections, the data products we adopt to retrieve \prot\ are KEPSEISMIC. For the analyses done in this work, a new correction to the long-term instrumental modulation was introduced and is  described in Appendix~\ref{appendix:data}. The long-term modulation has a periodicity consistent with the \kep\ orbital period, but it also produces significant signatures at different timescales, namely at the harmonics of the fundamental period. Another difference from the previous KOI studies is the rotation diagnostics applied in this work, which include the ACF, but also the wavelet analysis and the ACF-wavelet composite spectrum. Furthermore, here we only consider confirmed planet hosts and single-planet systems (to the best of our knowledge), and finally disregard possible binary and evolved stars, according to {\it Gaia} (Appendix~\ref{appendix:sample_selection}). Therefore, in addition to the data and rotation diagnostics, the target samples of this study and the previous studies differ. 

\begin{figure*}[h]
    \centering
    \includegraphics[width=\hsize]{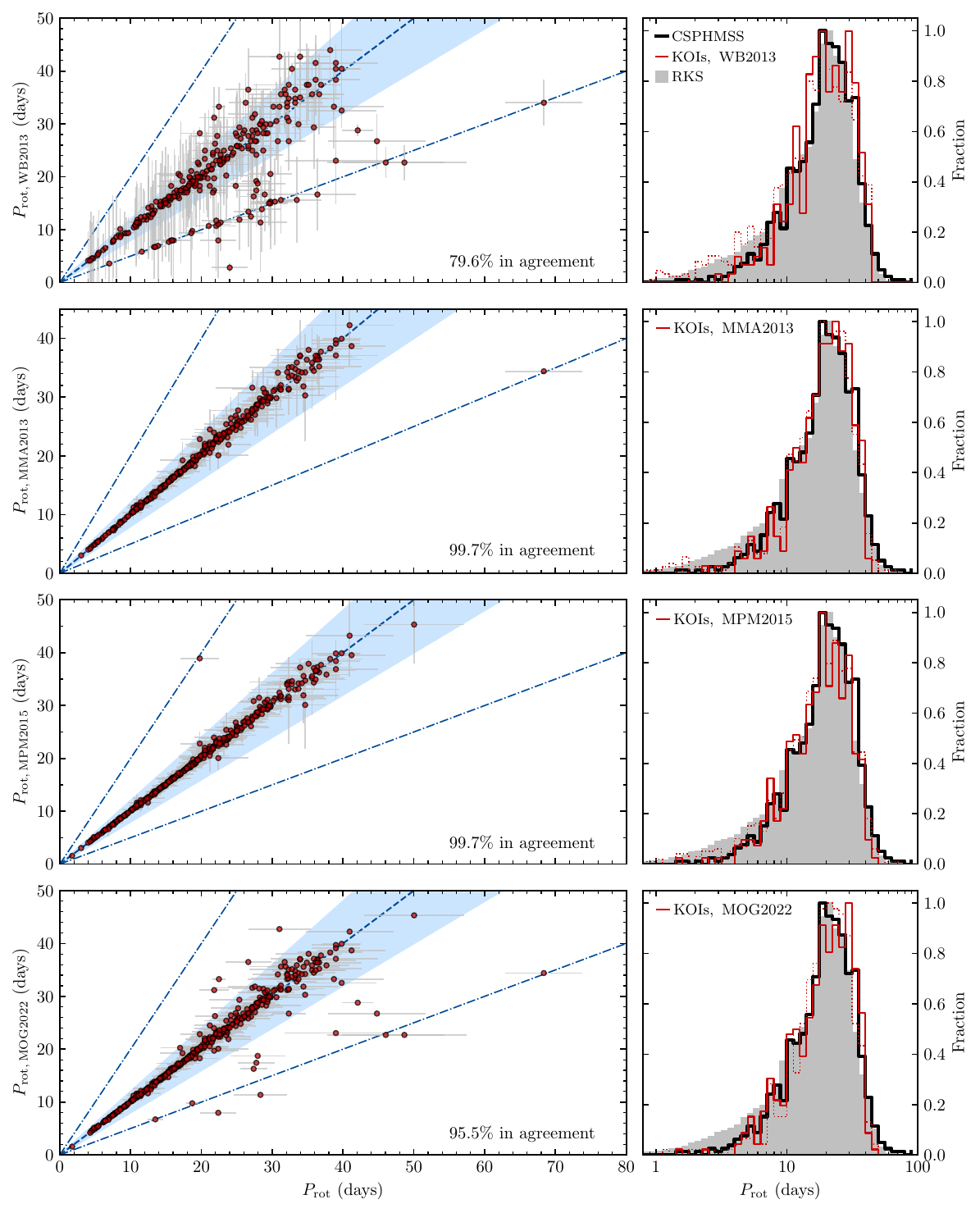}
    \caption{Comparison between the \prot\ estimates from the literature and from this work. {\it Left:} \prot\ from the literature as a function of   \prot\ from this work. The blue dashed and dot-dashed lines are the 1-1, 2-1, and 1-2 lines. In each panel   the percentage of common targets in agreement is indicated. The \prot\ estimates are considered to be in agreement if they are within 20\% (blue shaded region). {\it Right:} Comparison between the \prot distributions for the RKS (shaded grey region); CSPHMSS (thick black line); and the KOIs in the literature (red line). The solid red line shows the distributions only for the confirmed planet hosts, with only one known exoplanet, in the main sequence and single. The dotted histograms show the distributions for the original samples in each study (all system types considered).}
    \label{fig:Prot_comparison_hist}
\end{figure*}

For the comparison below we only keep the confirmed planet hosts with one known planet, and we remove evolved and binary stars from each sample, which reduces significantly the number of stars from each catalogue. The sample sizes after the selection cuts are summarized in Table~\ref{tab:koi_catalogs}. For the stars in common we have very good agreement for the \prot\ estimate (left panels of Fig.~\ref{fig:Prot_comparison_hist}) and we made sure also by visual inspection, as mentioned above, that we are reporting the correct rotation period. The best agreement with our work  was found with MMA2013 and MPM2015, which is expected given that we previously found an excellent agreement between the \prot\ estimates from our methodology and from that applied in MMA2013 and MPM2015 for the common targets \citep{2019ApJS..244...21S,2021ApJS..255...17S}. However, as we were performing our rotational analysis and comparing the results with MPM2015, we found that many of the targets for which we were able to measure \prot\ were flagged as non-reliable by MPM2015.  Many retrieved periods by MPM2015 correspond to instrumental artefacts; the MPM2015 periods are systematically longer than those retrieved by our analysis. 

\begin{table}[h]
    \caption{Sample sizes after applying the selection criteria described in Appendix~\ref{appendix:sample_selection} and only retaining the confirmed hosts with a (known) single planet.}
    \centering
    \resizebox{0.49\textwidth}{!}{%
    \begin{tabular}{c|ccc|cc}
         & \multicolumn{2}{c}{N stars} & \prot & \multicolumn{2}{c}{KS test}\\
         & all & common & agree & $D_\text{KS}$ & p-value \\\hline\hline
          WB2013 & 293 & 269 & 214 (79.5$\%$) & 0.05 & 0.73 \\
         MMA2013 & 322 & 312 & 311 (99.7 $\%$) & 0.04 & 0.77 \\
         MPM2015 & 374 & 365 & 364 (99.7$\%$) & 0.07 & 0.19 \\
         MOG2022 & 447 & 420 & 401 (95.5 $\%$) & 0.04 & 0.74 \\\hline
         RKS & 34,265 & -- & -- & 0.09 & $3\times10^{-6}$ \\\hline\hline
    \end{tabular}
    }
    \tablefoot{The first column lists the number of targets in each catalogue that survived the selection criteria, while the second column lists, from those, the number of targets in common with our CSPHMSS with \prot. The third column lists the number of targets with period estimates in agreement. The values in parentheses are the percentages with respect to the common stars. The last two columns summarize the results from the KS test, where $D_\text{KS}$ is the KS statistics.}
    \label{tab:koi_catalogs}
\end{table}

The right  panels of Fig.~\ref{fig:Prot_comparison_hist} compare the \prot\ distributions from this work and from the literature. The \prot\ distributions for the planet hosts are very similar. To quantify this agreement, we apply the KS test between our values and those of the literature; the results are summarized in Table~\ref{tab:koi_catalogs}. The KS test results do not reject the null hypothesis (p-value $>0.05$) that the samples come from the same distribution (for the KOIs). All the clean KOIs samples show a deficit of rotators in comparison to the RKS. Similarly to the results for the individual spectral types, the null hypothesis is rejected in the comparison between the KOIs and the RKS. For reference, we also plot the original distribution of reliable \prot\ from each study (WB2013, MMA2013, MPM2015, and MOG2022) without applying any selection criteria.

\FloatBarrier



\section{Estimate of the dead-zone radius in disks}
\label{app:rdz}

In this section we provide the derivation used to estimate the inner radius of the dead zone ($R_{\rm dz}$) in a stellar disk. This radius is then used as a lower limit for the initial semi-major axis of planets within ESPEM, which improves the initial conditions considered in the ESPEM population synthesis developed in \citetalias{2021A&A...650A.126A}.
For this estimation of $R_{\rm dz}$, we   suppose that the disk is passive ({i.e.} that it is heated only by the central star). The model derived here is based partially on the work of \citet{chiang_spectral_1997}. 

The first constraint on the radius of the dead zone comes from the criterion for the magneto-rotational instability (MRI). In the dead zone we consider the MRI to be inactive, which translates into
\begin{equation}
    \label{eq:MRI_crit}
    R_{m} = \frac{c_s^2}{\Omega_K\eta} < 4\pi^2\, ,
\end{equation}
where $R_m$ is the magnetic Reynolds number, $c_s$ is the speed of sound, $\Omega_K$ the \emph{Kepler}ian rotation rate, and $\eta$ the Ohmic dissipation coefficient \citep{2000prpl.conf..589S}. The speed of sound is related to the height of the disk $H$ and the local rotation rate ($\Omega_K$), and this gives

\begin{equation}
    \label{eq:eta_crit}
    \eta > \frac{H^2 \Omega_K}{4\pi^2}\, .
\end{equation}This can be recast as
\begin{equation}
    \label{eq:eta_crit_v2}
    \eta > 10^{18} \sqrt{\frac{M_\star}{\rm{M}_\odot}}\sqrt{\frac{R}{1 \mbox{ AU}}} \alpha_H^2\, ,
\end{equation}
where $\alpha_H = H/R$ is the disk thickness at radius $R$, $R$ is expressed in astronomical units, and $\eta$ is given in cm$^2$/s.
 
The ohmic dissipation in a disk can be classically estimated as a function of the temperature through the formula
\begin{equation}
    \label{eq:eta_vs_t}
    \eta = 4\times 10^6 T^{-1/2} {\rm exp}\left(25188/T \right)\, ,
\end{equation}
with $T$ the temperature of the disk in K and $\eta$ again in cm$^2$/s.
In the passive disk model of \citet{chiang_spectral_1997}, the temperature of the disk can be approximated in the stationary state to
\begin{equation}
    \label{eq:T_passive_disk}
    T = \frac{1}{2} \alpha_H^{1/4} \left(\frac{R}{R_\star}\right)^{-1/2} T_{\rm eff}\, ,
\end{equation}
with $T_{\rm eff}$ the stellar effective temperature.

Combining Equations \ref{eq:eta_crit_v2}, \ref{eq:eta_vs_t}, and \ref{eq:T_passive_disk} we obtain for the inner dead-zone radius $R_{\rm dz}$
\begin{equation}
    \label{eq:final_r_dz}
    \left(\frac{R_{\rm dz}}{R_\star}\right)^{-1/4}{\rm exp}\left( \frac{5\times 10^4}{T_\star} \left(\frac{R_{\rm dz}}{R_\star}\right)^{1/2} \alpha_H^{-1/4} \right) > 1.2\times 10^{10} \alpha_H^{17/8} T_\star^{1/2} \sqrt{\frac{M_\star}{\rm{M}_\odot}}\, ,
\end{equation}
where $T_{\rm eff}$ is expressed in Kelvin. The solution to this equation can be numerically fitted for solar-like stars ($M_\star \le 1.2 \rm{M}_\odot$), assuming a canonical value $\alpha_H=0.3$ to obtain the generic formulation
\begin{equation}
    R_{\rm dz} \simeq 3.42 - 1.47 \left(\frac{M_\star}{\rm{M}_\odot}\right) + 3.17 \left(\frac{M_\star}{\rm{M}_\odot}\right)^2 \,\, [10^{-2} {\rm AU}]\, .
\end{equation}

\section{ESPEM population considering only tidal interactions}
\label{sec:ESPEM_tideonly}

\begin{figure*}[!htb]
    \centering
    \includegraphics[width=0.76\textwidth]{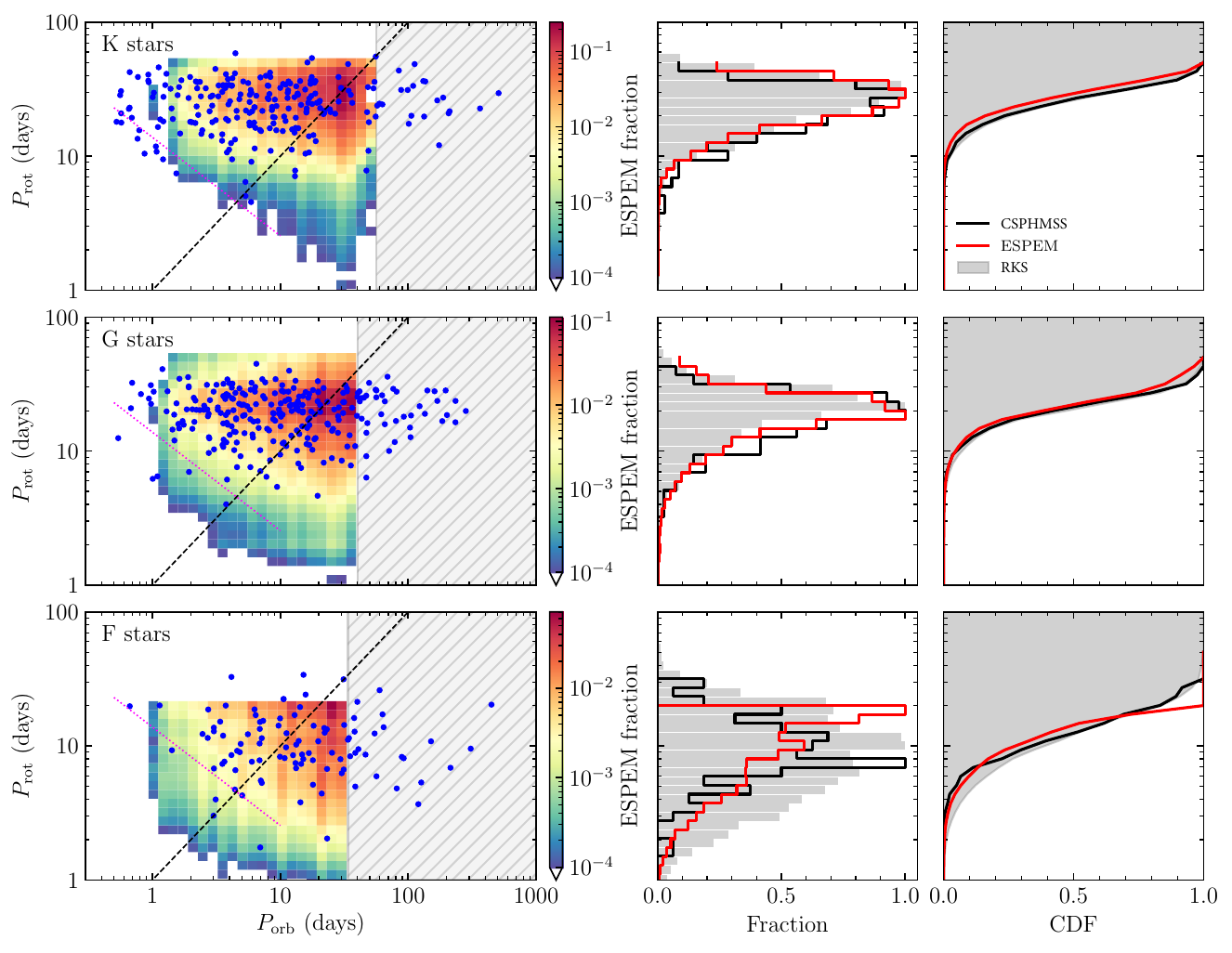}
    \caption{\prot\ vs \porb\ per spectral type. The layout is the same as Fig. \ref{fig:Prot_Porb_spectraltype}, but the ESPEM set shown considers only tidal interactions between the star and the planet.}
    \label{fig:Prot_Porb_spectraltype_tidleonly}
\end{figure*}

We show in Fig. \ref{fig:Prot_Porb_spectraltype_tidleonly} the same information as in Fig. \ref{fig:Prot_Porb_spectraltype}, but this time considering the ESPEM set of models where only tidal interactions are considered between the star and the planet. We highlight that in this case the model predicts even fewer planets on short-period orbits for all spectral types, showing that the additional migration of these planets is due to the magnetic torque they receive along their orbit. To date, the ESPEM set including magnetic torques agrees better with the observational population than the ESPEM set presented in Fig. \ref{fig:Prot_Porb_spectraltype_tidleonly}.

\FloatBarrier

\newpage

\section{Influence of   mass-loss rate and magnetic field scaling laws}
\label{app:laws}

\begin{figure*}[!htb]
    \centering
    \includegraphics[width=0.76\textwidth]{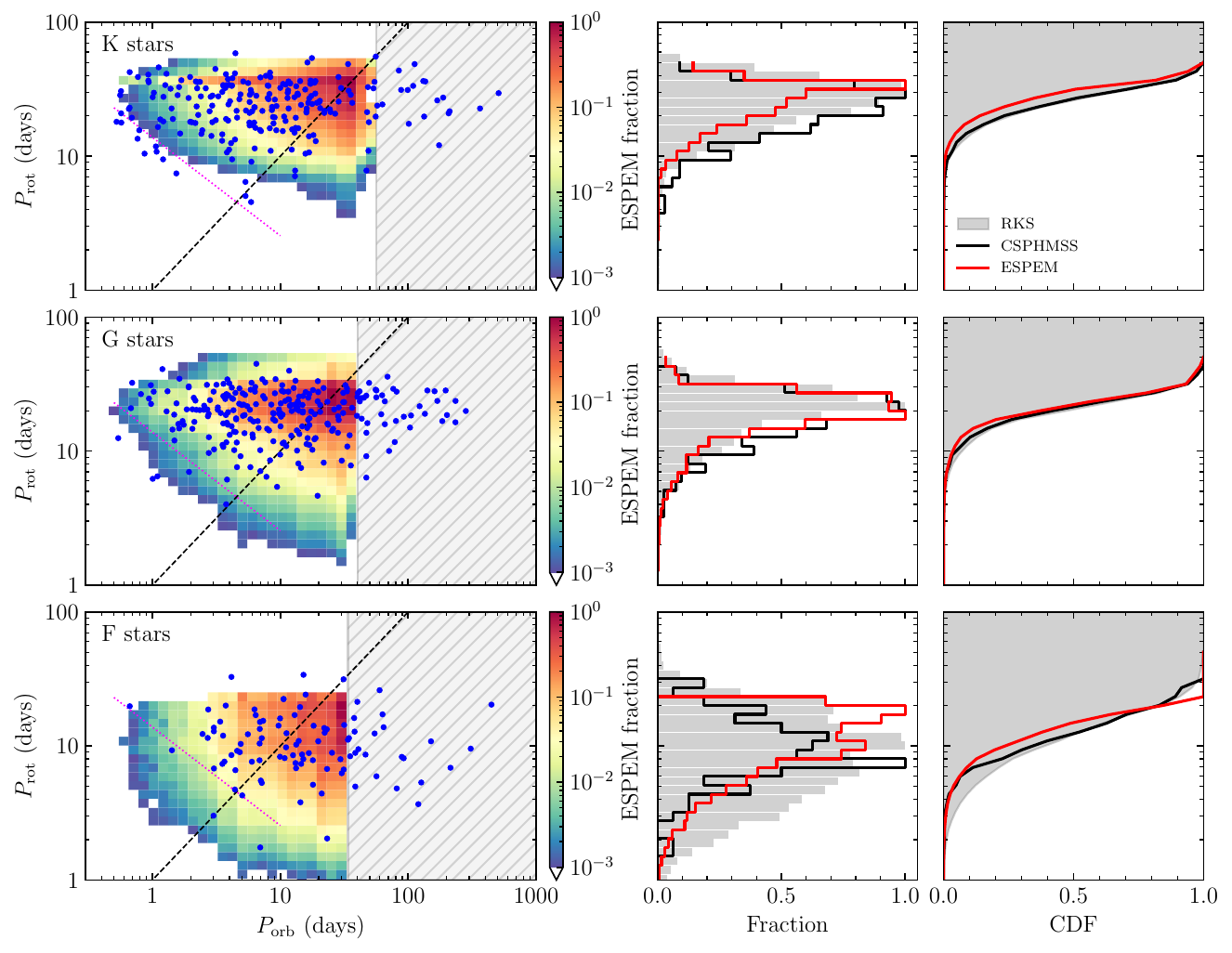}
    \caption{\prot\ vs \porb\ per spectral type. The layout is the same as Fig. \ref{fig:Prot_Porb_spectraltype}, but the ESPEM set shown was generated considering the scaling laws (\ref{eq:scaling_2}) for the evolution of the mass-loss rate and large-scale magnetic field of the star.}
    \label{fig:Prot_Porb_scaling_law_2}
\end{figure*}

Following \citet{2020A&A...635A.170A}, we   considered two different pairs of scaling laws to prescribe the evolution of the mass-loss rate and of the large-scale magnetic field of the stars modelled in this work. The two pairs of scaling laws are
\begin{equation}
\label{eq:scaling_1}
\left\{
\begin{aligned}
&B_\star [G] = 4 \left(\frac{Ro}{\rm{Ro}_\odot}\right)^{-1} \left(\frac{M_\star}{\rm{M}_\odot}\right)^{-1.76} \\
&\dot{M} [10^{-14} \rm{M}_\odot \mbox{yr}^{-1}] = 0.79 \left(\frac{\it{Ro}}{\rm{Ro}_\odot}\right)^{-2} \left(\frac{\it{M_\star}}{\rm{M}_\odot}\right)^{4}
\end{aligned}
\right. \, 
\end{equation}
and 
\begin{equation}
\label{eq:scaling_2}
\left\{
\begin{aligned}
&B_\star [G] = 4 \left(\frac{Ro}{\rm{Ro}_\odot}\right)^{-1.65} \left(\frac{M_\star}{\rm{M}_\odot}\right)^{-1.04} \\
&\dot{M} [10^{-14} \rm{M}_\odot \mbox{yr}^{-1}] = 0.79 \left(\frac{\it{Ro}}{{\rm{Ro}_\odot}}\right)^{-1} \left(\frac{\it{M_\star}}{\rm{M}_\odot}\right)^{2.9}
\end{aligned}
\right. \, ,
\end{equation}
where $Ro$ is the stellar Rossby number, as defined by \citet{2011ApJ...741...54C}. The two scaling laws were derived in \citet{2020A&A...635A.170A} to bracket the stellar evolution prescription under the available observational constraints on stellar rotation, wind, and magnetism. The population synthesis shown in Fig. \ref{fig:Prot_Porb_spectraltype} uses scaling laws (Equation~\ref{eq:scaling_1}). For the sake of completeness, we   also synthesized another population using scaling laws (Equation~\ref{eq:scaling_2}) to assess the dependency of our results on a specific choice of scaling law. The resulting population is shown in Fig. \ref{fig:Prot_Porb_scaling_law_2}. As expected, the change in the stellar braking law influences the distribution of stellar rotation rate. Because both braking laws are calibrated to the Sun, G-type stars (middle panel) show almost no differences between the two laws. Conversely, F-type stars seem to be better reproduced with the second scaling laws (Equation \ref{eq:scaling_2}), whereas K-type stars are not as well reproduced as the  experiment using the scaling laws (Equation \ref{eq:scaling_1}). Despite these differences, the conclusions of our study here still hold for both scaling laws: the ESPEM model systematically predicts a dearth in qualitative agreement with \emph{Kepler} CSPHMSS with a clear dependency on the spectral type. The exact dependency, though, can depend on the scaling laws considered for the braking torque.

\end{appendix}

\end{document}